\documentclass[aps,pra,showpacs,reprint,pacs]{revtex4-1}
\usepackage{float}
\usepackage{graphicx}% Include figure files
\usepackage{textcomp}
\usepackage{braket}
\usepackage{enumerate}
\usepackage{multirow}
\usepackage{array}
\usepackage{tabularx}
\usepackage{upgreek}
\usepackage{amsmath}

\begin{document}
\title{Study of CPO resonances on the intercombination line in $^{173}$Yb}
	
\author{Pushpander Kumar}
\author{Alok K. Singh}
\altaffiliation{Present address: Laboratoire Aim\'e Cotton, CNRS, Univ. Paris-Sud, ENS Cachan, B\^at. 505, 91405 Orsay, France.}
\author{Vineet Bharti}
\author{Vasant Natarajan}
\email{vasant@physics.iisc.ernet.in}	
\affiliation{Department of Physics, Indian Institute of Science, Bangalore-560012, India}
\author{Kanhaiya Pandey}
\affiliation{Department of Physics, Indian Institute of Technology, Guwahati 781039, India}

\begin{abstract}
We study coherent population oscillations (CPO) in an odd isotope of the two-electron atom Yb. The experiments are done using magnetic sublevels of the $ F_g = 5/2 \rightarrow F_e = 3/2 $ hyperfine transition in $^{173}$Yb of the $ {\rm {^1S_0} \rightarrow {^3P_1}} $ intercombination line. The experiments are done both with and without an appied magnetic field. In the absence of an applied field, the complicated sublevel structure along with the saturated fluorescence effect causes the linewidth to be larger than the 190 kHz natural linewidth of the transition. In the presence of a field (of magnitude 330 mG), a well-defined quantization axis is present which results in the formation of two M-type systems. The total fluorescence is then limited by spin coherence among the ground sublevels. In addition, the pump beam gets detuned from resonance which results in a reduced scattering rate from the $ {\rm ^3P_1} $ state. Both of these effects result in a reduction of the linewidth to a subnatural value of about 100 kHz.
\end{abstract}
\maketitle	

\section{Introduction}
Coherent population oscillation (CPO) is a phenomenon in which two levels are coupled by two coherent electromagnetic fields of different amplitudes and frequencies \cite{BLB03}. This results in the population difference between the two levels to oscillate periodically at the beat frequency of the two fields. The linewidth of the resulting resonance is limited by the population relaxation rate, and can be subnatural for optical transitions. In recent times, it has received much interest from the quantum optics community \cite{LKB12, KLA13, MBG14}, because narrow CPO resonances can be used for applications in light storage \cite{ASM14,NMB17} and optical memories \cite{EAS10}. It has other applications similar to the related phenomenon of coherent population trapping (CPT) \cite{ARI96}, but is more advantageous because of the following reasons.
\begin{enumerate}[(i)]
\item There is no third level involved, so it can be observed in a larger class of materials including gases, liquids, and room temperature solids.

\item It does not require a two-photon resonance, therefore laser jitter---both frequency and amplitude---becomes irrelevant. 

\item It offers spectral features that are broad, which implies that the medium will respond to very short optical pulses (of picosecond duration or less), whereas the shortest pulses that can be manipulated by CPT-based schemes are of nanosecond duration.
\end{enumerate} 

We have recently reported the first experimental observation of CPT and CPO in the two-electron atom Yb \cite{SIN15}, a theoretical model for which was presented in Ref.\ \cite{VMA16}. The study was novel because it was an observation in a V-type system, a system which does not have such resonances in a one-electron atom. We used the intercombination line ($\rm ^1S_0 \rightarrow {^3P_1}$  transition) at 556 nm and the even isotope $^{174}{\rm Yb}$.  We used a small magnetic field (of order 0.5 G) to shift the CPT resonances away from line center, and retain only the CPO resonances. The linewidth of the central peak was then about 400 kHz, which was a factor of two higher than the natural linewidth of the upper $\rm ^3P_1$ state. This was because the peak was due to multiple CPO resonances, each of which had a resonant pump field.

Here, we report a study of CPO resonances in the odd isotope $^{173}$Yb. The main difference from our work on the even isotope is that there is a single peak with and without a magnetic field. This is because there are only CPO resonances and no CPT resonances. CPT resonances will move in the presence of a magnetic field, because the field will cause the pump beam to become detuned and the resonance will appear at the location where the probe beam has the same detuning. For CPO resonances, such detuning does not cause any shift from line center because it only deals with the two levels being coupled, but will cause a reduction in decoherence due to spontaneous emission from the excited state. The reduced scattering rate will result in a reduction in the linewidth of the CPO peak. The detuning also causes a decrease in the fluorescence signal, and an overall decrease in the signal-to-noise ratio (SNR). This explanation is borne out by a detailed density-matrix calculation of the levels involved.

We demonstrate this using the $ F_g = 5/2 \rightarrow F_e = 3/2 $ hyperfine transition  $^{173}$Yb, but the same trend of linewidth reduction is observed for the other two hyperfine transitions in this isotope. $\rm{^{173}Yb} $ was preferred over the other odd isotope $ \rm{^{171}Yb} $---which has only one transition---for the following reason. It gives us the ability to study transitions where the number of sublevels in the excited state is of three different kinds: smaller ($ 5/2 \rightarrow 3/2 $), equal ($ 5/2 \rightarrow 5/2 $), and larger ($ 5/2 \rightarrow 7/2 $). 

This ability to study transitions with different number of sublevels is one of the main differences from the earlier work on CPO resonances done with metastable He in Ref.~\cite{LKB12}. In addition, those experiments were done with a vapor cell inside a magnetic shield. By contrast, our experiments are done with an atomic beam, because a vapor cell gets coated with Yb and becomes opaque due to the high sublimation temperature of Yb (about 500\textdegree C). Therefore, it is practically impossible to make a vapor cell of Yb; hence all our previous Yb work has been done using fluorescence detection from an atomic beam \cite{PSK09,SAN13}. The theoretical model given in Ref.~\cite{LKB12} is also not applicable to our case mainly because of the large number of levels involved.

\section{Experimental details}
\begin{figure}
	\centering
	\includegraphics[width=0.95\linewidth]{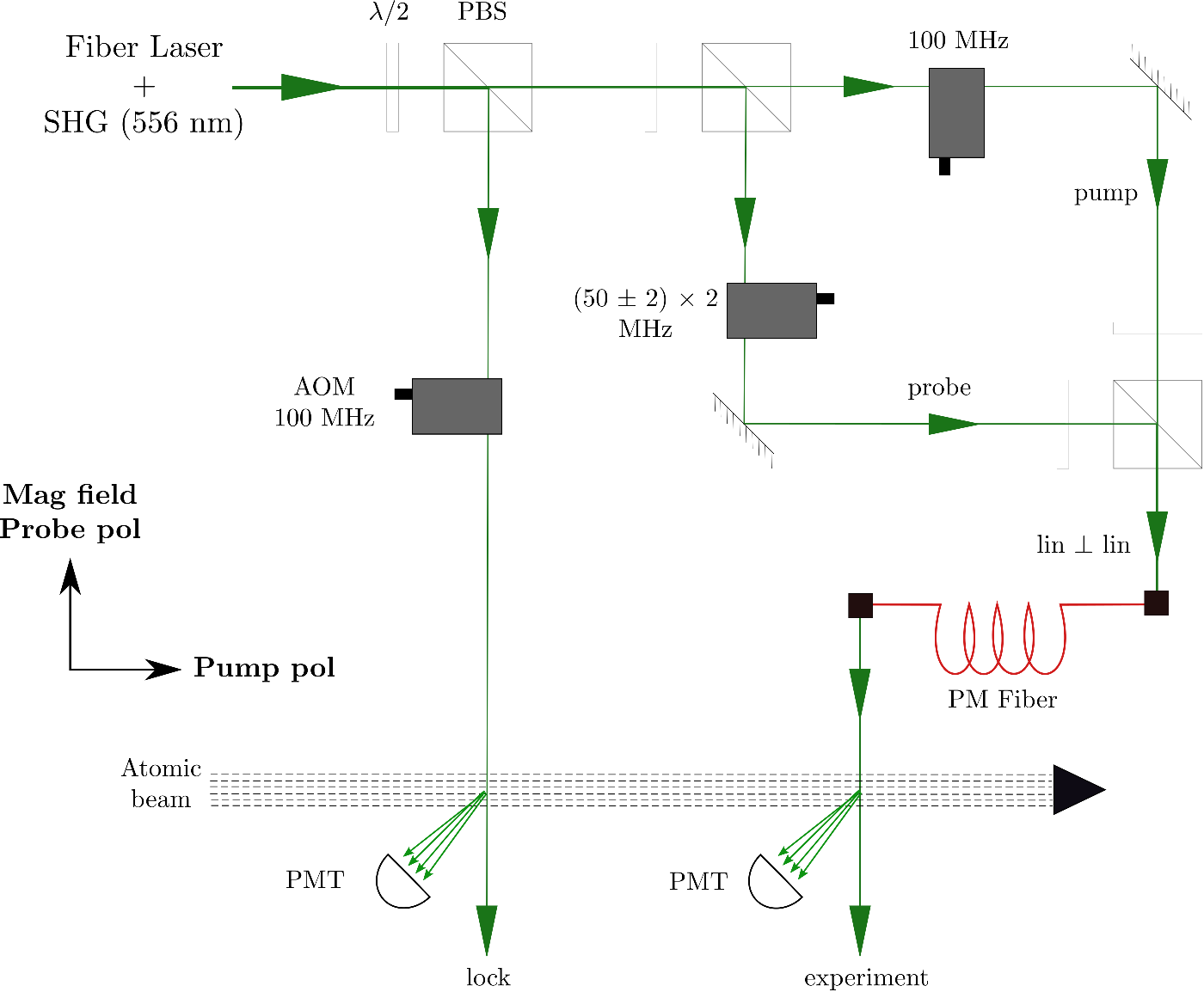}
	\caption{Experimental setup. The probe beam is scanned by $\pm 4$ MHz using a double-passed AOM. Figure key: $\lambda$/2---halfwave retardation plate; PBS---polarizing beam splitter; PM fiber---polarizing maintaining fiber; PMT---photomultiplier tube.}
	\label{schematic}
\end{figure}

The experimental setup shown in Fig.\ \ref{schematic} is the same as in our previous work in Ref.\ \cite{SIN15}, and is reviewed here for completeness. Briefly, the $\rm ^1S_0 \rightarrow {^3P_1} $ transition at 556 nm is generated by doubling the output of a fiber laser with linewidth of 70 kHz. The doubling is done in a commercial external cavity doubler using potassium niobate as the nonlinear crystal. As seen in the figure, part of the output beam from the doubler is used to lock the laser to the correct hyperfine transition. To achieve this, the laser beam is sent perpendicular to the atomic beam, and the resulting fluorescence is collected using a photomultiplier tube (PMT). The frequency dither required for generating the error signal is obtained by frequency modulating the acousto-optic modulator (AOM) in the path of the laser beam.

The remaining part of the laser beam is used for the experiment. It is divided into two parts---one as a pump beam and the other as a probe beam---with orthogonal linear polarizations using a polarizing beam splitter (PBS). The pump beam is fixed in frequency, and is resonant with the transition used for locking the laser. This is achieved by using a second AOM with an identical shift (of $ +100 $ MHz) as the one used for laser locking. The probe beam is scanned around this transition by double passing a third AOM at $50\pm 2 $ MHz. The double passing ensures directional stability as the AOM is scanned, and the variable shift of $ \pm 2 $ MHz allows the probe beam to be scanned by $\pm 4$ MHz around the resonance. The two beams are combined on another PBS, and then transported to the experimental chamber using a polarization maintaining (PM) fiber. The use of the fiber ensures perfect overlap of the two beams as the probe beam is scanned. The mode coming out of the fiber is perfectly Gaussian. The power in each beam is 40 $\upmu$W and its $ 1/e^2$ diameter is 5 mm, which gives a maximum Rabi frequency of $ 1.2 \, \Gamma $ at beam center. The output of the fiber is sent perpendicular to the atomic beam, and the resulting fluorescence signal is detected using a second identical PMT.

The atomic beam used both for laser locking and the experiment is contained in an ultra high vacuum (UHV) chamber. The beam is generated by resistively heating a quartz ampule containing an ingot of unenriched Yb to a temperature of about $ 400^\circ $C. The vacuum chamber is maintained at a pressure below $10^{-8}$ torr using a 20 l/s ion pump.

The magnetic field required for the experiment is generated by winding three pairs of Helmholtz coils in three orthogonal directions on the outside of the UHV chamber. The value of the B field was measured with a three-axis fluxgate magnetometer. The first set of experiments was done with nominally zero field. However, the presence of stray fields in the lab, coupled with the fact that no magnetic shield was used, meant that there was always a small non-zero field. The second set of experiments was done with a finite longitudinal field, which was generated by increasing the current in the pair of coils that is coaxial with the laser beam. The value of 330 mG is the same as the one used in our even isotope work, and it was chosen so that CPT resonances move significantly away from line center in the presence of the field. The stability of the power supply driving the coils is better than $ 10^{-4}$. 

\section{Experimental results}

\begin{figure}
	\centering
	\includegraphics[width=0.8\linewidth]{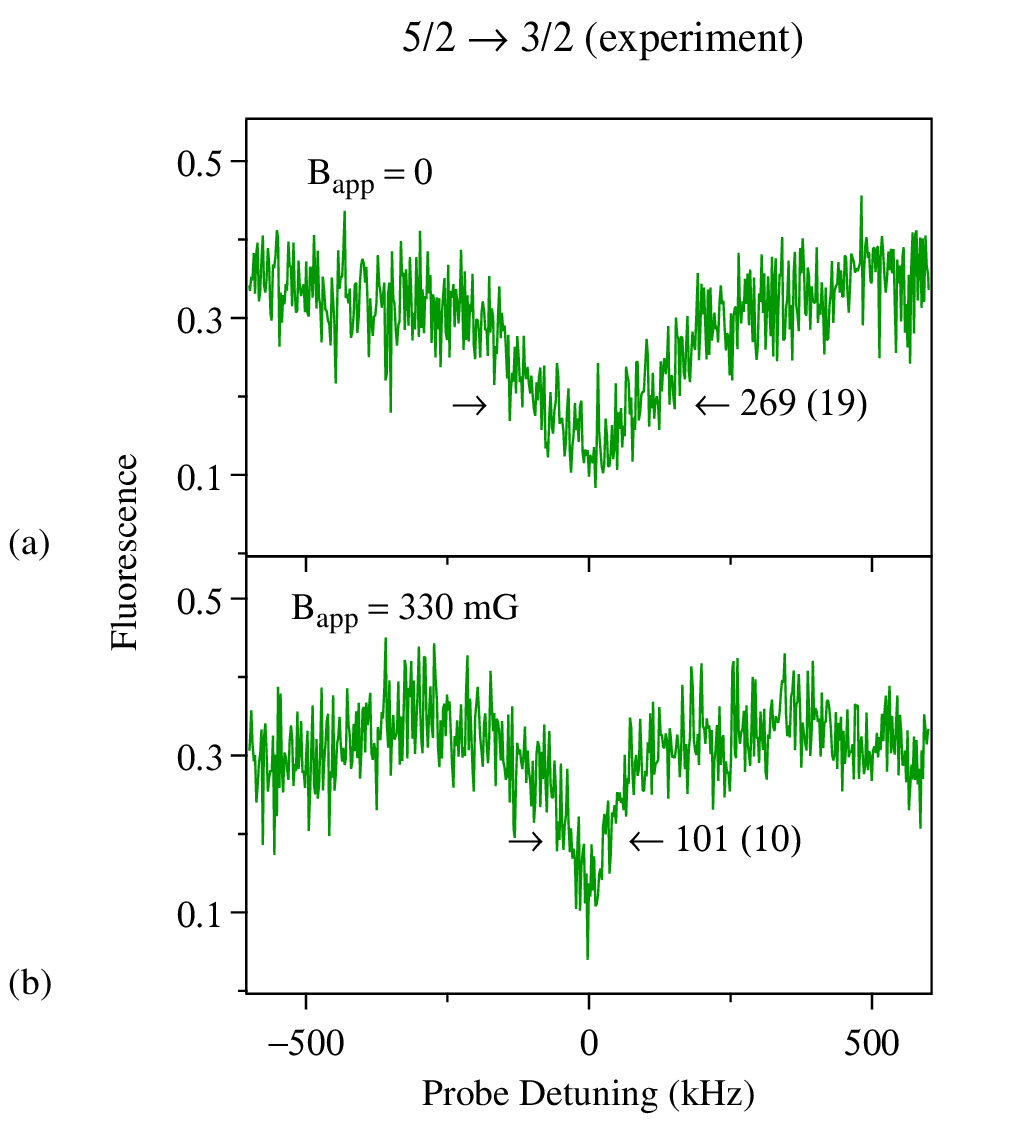}
	\caption{Experimental spectra for the $ 5/2 \rightarrow 3/2 $ hyperfine transition, showing a single peak due to multiple CPO and CPT resonances. (a) With no applied B field. (b) With an applied B field of 330 mG.}
	\label{3/2-5/2}
\end{figure}

$ ^{173}{\rm Yb} $ has a nuclear spin $ I $ of $ 5/2 $, hence there are three hyperfine transitions in this line: $ 5/2 \rightarrow 3/2 $, $ 5/2 \rightarrow 5/2 $, and $ 5/2 \rightarrow 7/2 $. Experimental spectra for the first transition are shown in Fig.\ \ref{3/2-5/2}. The first thing to notice is that there is a single peak, both with and without an applied B field. The linewidth of the peak without a field---but with a resonant pump beam---is 270 kHz. This is slightly larger than the natural linewidth of the intercombination line of 190 kHz (determined by the lifetime of the $\rm ^3P_1 $ state), and arises both because of multiple CPO resonances contributing to the peak and a saturated fluorescence effect. The saturated fluorescence effect is a two-level process in which there is a dip in the fluorescence at zero detuning. This has been previously observed in the intercombination line in Sr \cite{YPP15}. In the presence of a B field of 330 mG, the linewidth reduces to a subnatural value of 100 kHz. This is primarily because of reduced spontaneous emission from the upper state---the B field causes the pump beam to be detuned---as shown in our work on CPT resonances \cite{KKB17}. While this is the main effect, other mechanisms limiting the linewidth include the ground-state decoherence rate, power broadening, and transit-time broadening. The relative contributions of these factors is considered in detail in the ``Theoretical Analysis'' section. 

\begin{figure}
	\centering
	\includegraphics[width=0.8\linewidth]{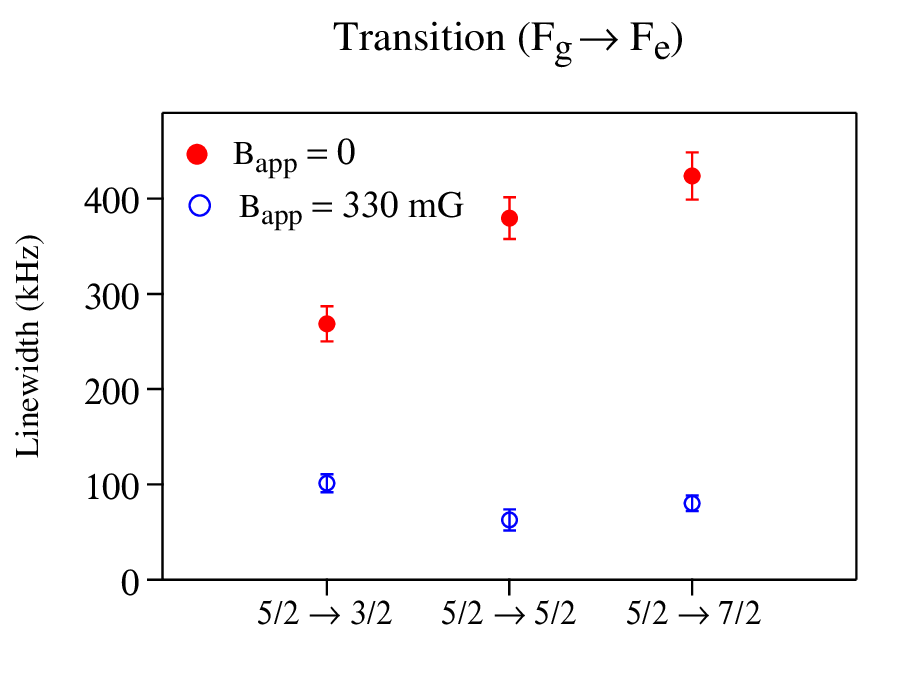}
	\caption{Linewidth of transitions without and with an applied B field. The error bars are the $ 1 \, \sigma $ deviations obtained from the Lorentzian fit.}
	\label{yb173linewidth}
\end{figure}

The same behavior, i.e.\ a linewidth reduction in the presence of a B field, is seen for the other two hyperfine transitions---$ 5/2 \rightarrow 5/2 $ and $ 5/2 \rightarrow 7/2 $. The results of linewidth measurements for all three cases are shown in Fig.\ \ref{yb173linewidth}. As seen, the linewidth of the resonance for the no-field case shows a steady increase from the one for the $ 5/2 \rightarrow 3/2 $ transition. This is because the number of sublevels increases steadily, which results in more and more decay channels contributing to the peak.

\section{Theoretical analysis}
\begin{figure}
	\centering
	\includegraphics[width=0.99\linewidth]{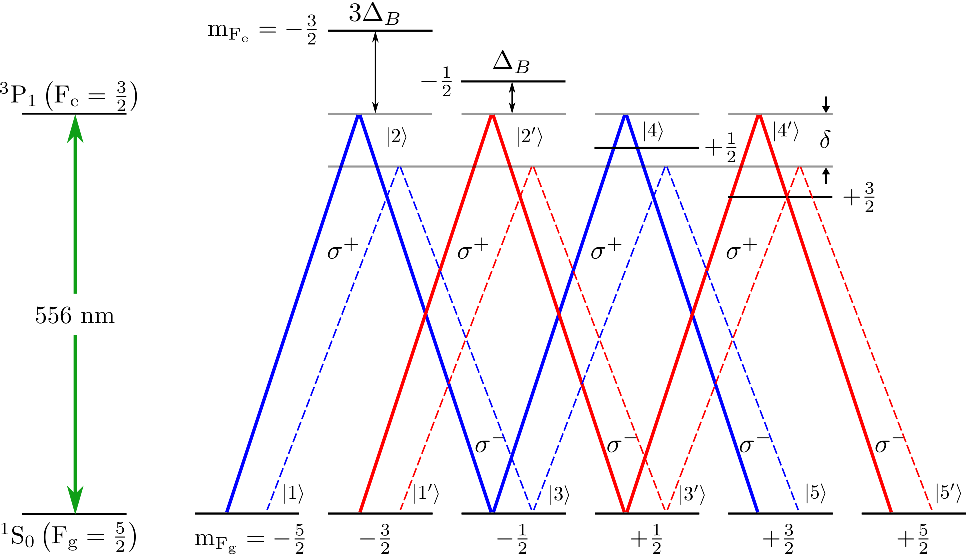}
	\caption{Magnetic sublevels of the $ 5/2 \rightarrow 3/2 $ hyperfine transition in $^{173}{\rm Yb} $. Transitions coupled by the pump beam are shown with solid lines, while those coupled by the probe beam are shown with dotted lines. The probe beam is scanned in frequency with a variable detuning $ \delta $. The pump beam has a fixed frequency resonant with the zero-field transition, and gets detuned in the presence of a B field. The magnitude of the shift for each excited sublevel is proportional to $ \Delta_B $.}
	\label{levels}
\end{figure}

The complete sublevel structure for the $ 5/2 \rightarrow 3/2 $ transition in the presence of a B field is shown in Fig.~\ref{levels}. The presence of the field defines a quantization axis, which is independent of the direction of light polarization. Therefore, even though the light beams are linearly polarized, we do not have to consider transitions coupled by $ \pi $ polarization (selection rule $\Delta m =0 $). Furthermore, the role of $ \pi $ polarization has been discussed in detail in our earlier work on the even isotope \cite{SIN15}. It is also reasonable to consider only transitions in the presence of a B field because, as mentioned in the experimental details section, it is practically impossible to nullify stray fields completely.

Each excited state sublevels shifts by an amount proportional to $ \Delta_B $, which is 139 kHz for our experimental field of 330 mG. The pump and probe beams are linearly polarized. This means that they decompose into equal amounts of right circular ($ \sigma^+ $) and left circular ($ \sigma^- $) polarizations. $ \sigma^+ $ couples transitions with the selection rule $ \Delta m = +1 $ and $ \sigma^- $ couples with selection rule $ \Delta m = -1 $.  

Transitions between sublevels coupled by the pump and probe beams depicted in Fig.~\ref{levels} show that the system separates into two M-type subsystems. This shows that no CPT-like resonances are formed, because CPT requires a simple three level $ \Lambda$-type system.

In order to see this theoretically, we do a density-matrix analysis of this system. To enable this we depict the two types of transition in the figure with different colors. The first one (shown in blue) involves the ground sublevels $ m_{F_g} = -5/2, -1/2, \text{and} +3/2 $. The second one (shown in red) involves sublevels $ m_{F_g} = -3/2, +1/2, \text{and} +5/2 $. For ease of writing the Hamiltonian, we define the levels in the first system from $ \ket{1} $ to $ \ket{5} $. The Rabi frequency of the probe beam driving the $ \ket{i} \rightarrow \ket{j} $ transition is $ \Omega^p_{ij} $ while its detuning is $ \delta^p_{ij} $. The corresponding parameters for the pump beam are $ \Omega^c_{ij} $ and $ \delta^c_{ij} $. The Rabi frequencies of the probe and pump beams are weighted by the respective Clebsch-Gordan coefficients. In the rotating wave approximation, the Hamiltonian is 
\begin{equation}
\begin{aligned}
H&=0\ket{1}\bra{1}\\
&-\hbar\delta^p_{12}\ket{2}\bra{2}\\
&-\hbar(\delta^p_{12}-\delta^p_{23})\ket{3}\bra{3} \\
&-\hbar(\delta^p_{12}-\delta^p_{23}+\delta^p_{34})\ket{4}\bra{4}\\
&-\hbar(\delta^p_{12}-\delta^p_{23}+\delta^p_{34}-\delta^p_{45})\ket{5}\bra{5}\\
&+\left(\frac{\hbar\Omega^p_{12}}{2}+\frac{\hbar\Omega^c_{12}}{2}e^{-i\hbar(\delta^p_{12}-\delta^c_{12})t}\right)\ket{1}\bra{2}\\
&+\left(\frac{\hbar\Omega^p_{23}}{2}+\frac{\hbar\Omega^c_{23}}{2}e^{i\hbar(\delta^p_{23}-\delta^c_{23})t}\right)\ket{2}\bra{3} \\
&+\left(\frac{\hbar\Omega^p_{34}}{2}+\frac{\hbar\Omega^c_{34}}{2}e^{-i\hbar(\delta^p_{34}-\delta^c_{34})t}\right)\ket{3}\bra{4}\\
&+\left(\frac{\hbar\Omega^p_{45}}{2}+\frac{\hbar\Omega^c_{45}}{2}e^{i\hbar(\delta^p_{45}-\delta^c_{45})t}\right)\ket{4}\bra{5}\\
&+\text{Hermitian conjugate}
\label{Hamiltonian}
\end{aligned}
\end{equation}
As seen from the figure, the probe detunings are such that 
\begin{equation}
\delta^p_{12} -\delta^p_{23} = 0 \quad \text{and} \quad \delta^p_{34}-\delta^p_{45} = 0
\end{equation}
since there is no splitting of the ground state. 
The equation of motion of the density matrix $ \rho $ is given by 
\begin{equation}
\dot{\rho} = -\dfrac{i}{\hbar}[H, \rho] - \dfrac{1}{2}\{\Gamma, \rho\}
\label{rhodot}
\end{equation}
where $ \Gamma $ is the decay matrix. Its diagonal terms give the total decay rate (radiative and non-radiative) of the respective population terms. The off-diagonal terms represent the decoherence rate between states $ \ket{i} $ and $ \ket{j} $, such that
\begin{equation}
\Gamma_{ij} = \dfrac{\Gamma_{ii} + \Gamma_{jj}}{2}
\end{equation}
$ \Gamma_{ij} $ between two ground states is the spin relaxation rate, and is taken to be $ 2\pi \times 10 $ kHz.

The solution of Eq.~\eqref{rhodot} will be time dependent because the Hamiltonian is time dependent. Hence the solution can be written in the form 
\begin{equation}
\rho_{ij}=\rho^0_{ij}+\rho^{\pm1}_{ij}e^{\pm i(\delta^p_{ij} - \delta^c_{ij})t}+\rho^{\pm2}_{ij}e^{\pm 2i(\delta^p_{ij} - \delta^c_{ij})t}+...
\end{equation}
$ \rho^0 $---the DC component of $ \rho $---is found by taking the Fourier transform of the numerical solution of $ \rho $, which includes all orders in the expansion. 

\begin{figure}
	\centering
	\includegraphics[width=0.8\linewidth]{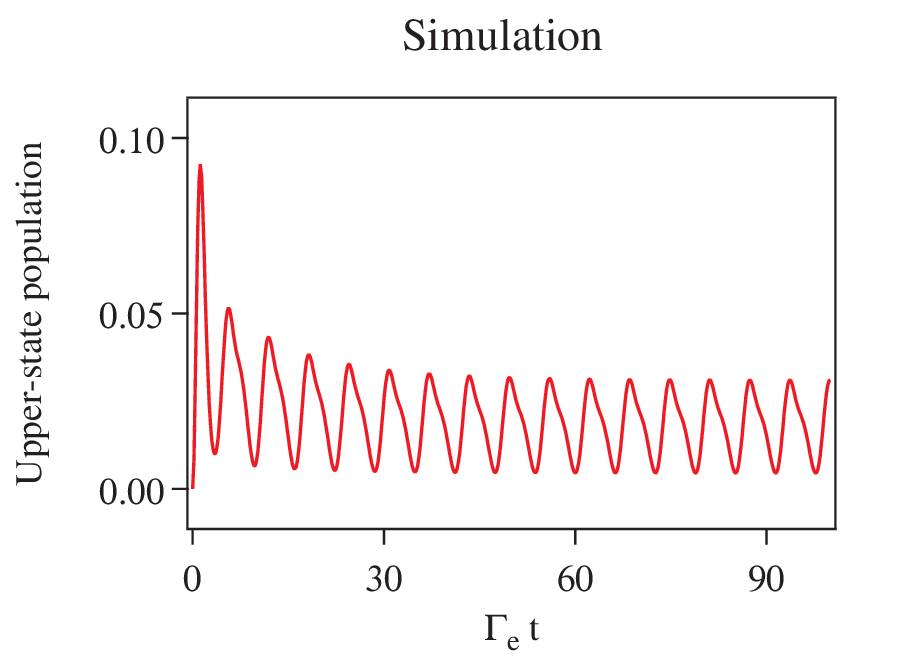}
	\caption{Transient behavior of $\rho_{22}+\rho_{44}$ (populations in upper levels) for the first M-type system shown in Fig.~\ref{levels}. $\Gamma_e$ is the spontaneous decay rate of the excited state. The value of the B field is taken as 330 mG, and the Rabi frequencies of both the beams are taken to be at the maximum value from the experiment (1.2 $\Gamma_e$).}
	\label{transientsim}
\end{figure}

The fluorescence signal obtained in the experiment is proportional to the population in the upper levels, i.e. $ \rho_{22} + \rho_{44} $. From the time dependence of this quantity shown in Fig.~\ref{transientsim}, we see that it oscillates around a mean value after a few lifetimes of the excited state. Thus, the system reaches ``steady state'' after this time.

\begin{figure}
	\centering
	\includegraphics[width=0.9\linewidth]{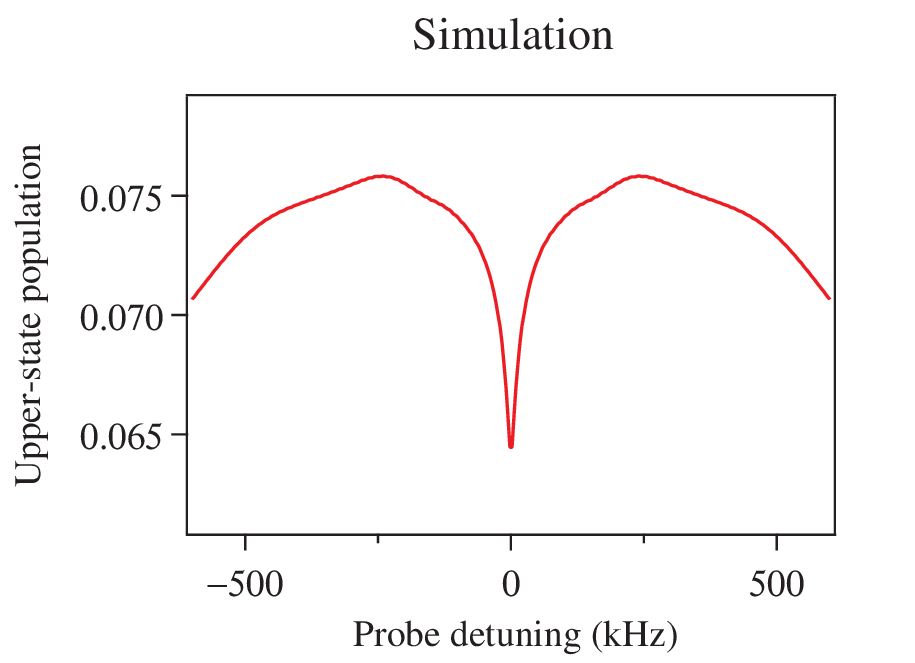}
	\caption{Populations in the upper levels as a function of probe detuning for the two M-type systems formed using magnetic sub-levels of the $ 5/2 \rightarrow 3/2 $ transition in the presence of B field of 330 mG. The Rabi frequencies of both the beams are taken to be at the maximum value from the experiment (1.2 $\Gamma_e$).}
	\label{cposim}
\end{figure}

The other M-type system is analyzed in a similar manner. For clarity, it is denoted with prime symbols in Fig.~\ref{levels}. The ``steady state'' populations in the upper levels for the two M-type systems, i.e. $ \rho_{22} + \rho_{44} + \rho_{2^{\prime}2^{\prime}} + \rho_{4^{\prime}4^{\prime}} $, are shown in Fig.~\ref{cposim}. \textit{The main thing to notice is that there is only one peak at line center, exactly as seen in the experimental spectrum}. This is different from the spectrum for the even isotope, shown in Ref.~\cite{SIN15}, where four shifted peaks corresponding to CPT resonances were seen in the presence of a B field. The linewidth of the peak is smaller than the natural linewidth because detuning will cause reduced decoherence from the upper state. But this comes at the price of reduced fluorescence signal, something which is also observed experimentally.

We have considered other factors limiting the linewidth. Ground state decoherence rate plays a negligible role. Transit-time broadening \cite{THQ80} contributes to less than 20\%. Power broadening \cite{MDP17} contributes to less than 50\%. Therefore, reduced scattering from the excited state remains the dominant factor.

The theoretical model presented above reproduces the fact that the experimentally observed linewidth is subnatural. But it does not give the exact linewidth observed, probably because it assumes a constant Rabi frequency across the laser beam while in reality it is a Gaussian distribution. 

\section{Conclusions}
In summary, we have studied CPO resonances in the odd isotope $\rm {^{173}Yb} $. The magnetic sublevel structure of the intercombination line used in this study is such that the peak at line center does not split into multiple peaks in the presence of a B field, which is quite different from the even isotope used in our previous work \cite{SIN15}. The effect of a B field is to cause the pump beam to be detuned from resonance, which results in a reduced scattering rate from the excited state of the transition. The effect of this on the CPO resonance peak is to reduce its linewidth. The linewidth reduces to a subnatural value of 100 kHz with a magnetic field of 330 mG. Such features could lead to new mechanisms for sub-Doppler cooling in divalent atoms compared to the more commonly used monovalent alkali-metal atoms \cite{MCF16}.

\section*{Acknowledgments}
This work was supported by the Department of Science and Technology, India. P K and A K S acknowledge financial support from the Council of Scientific and Industrial Research, India; and V B from a D S Kothari post-doctoral fellowship of the University Grants Commission, India. The authors thank S Raghuveer for help with manuscript preparation.

%\bibliography{eitrefs}

\begin{thebibliography}{17}%
	\makeatletter
	\providecommand \@ifxundefined [1]{%
		\@ifx{#1\undefined}
	}%
	\providecommand \@ifnum [1]{%
		\ifnum #1\expandafter \@firstoftwo
		\else \expandafter \@secondoftwo
		\fi
	}%
	\providecommand \@ifx [1]{%
		\ifx #1\expandafter \@firstoftwo
		\else \expandafter \@secondoftwo
		\fi
	}%
	\providecommand \natexlab [1]{#1}%
	\providecommand \enquote  [1]{``#1''}%
	\providecommand \bibnamefont  [1]{#1}%
	\providecommand \bibfnamefont [1]{#1}%
	\providecommand \citenamefont [1]{#1}%
	\providecommand \href@noop [0]{\@secondoftwo}%
	\providecommand \href [0]{\begingroup \@sanitize@url \@href}%
	\providecommand \@href[1]{\@@startlink{#1}\@@href}%
	\providecommand \@@href[1]{\endgroup#1\@@endlink}%
	\providecommand \@sanitize@url [0]{\catcode `\\12\catcode `\$12\catcode
		`\&12\catcode `\#12\catcode `\^12\catcode `\_12\catcode `\%12\relax}%
	\providecommand \@@startlink[1]{}%
	\providecommand \@@endlink[0]{}%
	\providecommand \url  [0]{\begingroup\@sanitize@url \@url }%
	\providecommand \@url [1]{\endgroup\@href {#1}{\urlprefix }}%
	\providecommand \urlprefix  [0]{URL }%
	\providecommand \Eprint [0]{\href }%
	\providecommand \doibase [0]{http://dx.doi.org/}%
	\providecommand \selectlanguage [0]{\@gobble}%
	\providecommand \bibinfo  [0]{\@secondoftwo}%
	\providecommand \bibfield  [0]{\@secondoftwo}%
	\providecommand \translation [1]{[#1]}%
	\providecommand \BibitemOpen [0]{}%
	\providecommand \bibitemStop [0]{}%
	\providecommand \bibitemNoStop [0]{.\EOS\space}%
	\providecommand \EOS [0]{\spacefactor3000\relax}%
	\providecommand \BibitemShut  [1]{\csname bibitem#1\endcsname}%
	\let\auto@bib@innerbib\@empty
	%</preamble>
	\bibitem [{\citenamefont {Bigelow}\ \emph {et~al.}(2003)\citenamefont
		{Bigelow}, \citenamefont {Lepeshkin},\ and\ \citenamefont {Boyd}}]{BLB03}%
	\BibitemOpen
	\bibfield  {author} {\bibinfo {author} {\bibfnamefont {M.~S.}\ \bibnamefont
			{Bigelow}}, \bibinfo {author} {\bibfnamefont {N.~N.}\ \bibnamefont
			{Lepeshkin}}, \ and\ \bibinfo {author} {\bibfnamefont {R.~W.}\ \bibnamefont
			{Boyd}},\ }\href {\doibase 10.1103/PhysRevLett.90.113903} {\bibfield
		{journal} {\bibinfo  {journal} {Phys. Rev. Lett.}\ }\textbf {\bibinfo
			{volume} {90}},\ \bibinfo {pages} {113903} (\bibinfo {year}
		{2003})}\BibitemShut {NoStop}%
	\bibitem [{\citenamefont {Laupr\^etre}\ \emph {et~al.}(2012)\citenamefont
		{Laupr\^etre}, \citenamefont {Kumar}, \citenamefont {Berger}, \citenamefont
		{Faoro}, \citenamefont {Ghosh}, \citenamefont {Bretenaker},\ and\
		\citenamefont {Goldfarb}}]{LKB12}%
	\BibitemOpen
	\bibfield  {author} {\bibinfo {author} {\bibfnamefont {T.}~\bibnamefont
			{Laupr\^etre}}, \bibinfo {author} {\bibfnamefont {S.}~\bibnamefont {Kumar}},
		\bibinfo {author} {\bibfnamefont {P.}~\bibnamefont {Berger}}, \bibinfo
		{author} {\bibfnamefont {R.}~\bibnamefont {Faoro}}, \bibinfo {author}
		{\bibfnamefont {R.}~\bibnamefont {Ghosh}}, \bibinfo {author} {\bibfnamefont
			{F.}~\bibnamefont {Bretenaker}}, \ and\ \bibinfo {author} {\bibfnamefont
			{F.}~\bibnamefont {Goldfarb}},\ }\href {\doibase 10.1103/PhysRevA.85.051805}
	{\bibfield  {journal} {\bibinfo  {journal} {Phys. Rev. A}\ }\textbf {\bibinfo
			{volume} {85}},\ \bibinfo {pages} {051805} (\bibinfo {year}
		{2012})}\BibitemShut {NoStop}%
	\bibitem [{\citenamefont {Kapale}\ \emph {et~al.}(2013)\citenamefont {Kapale},
		\citenamefont {Lam},\ and\ \citenamefont {Agarwal}}]{KLA13}%
	\BibitemOpen
	\bibfield  {author} {\bibinfo {author} {\bibfnamefont {K.~T.}\ \bibnamefont
			{Kapale}}, \bibinfo {author} {\bibfnamefont {W.~K.}\ \bibnamefont {Lam}}, \
		and\ \bibinfo {author} {\bibfnamefont {G.~S.}\ \bibnamefont {Agarwal}},\ }in\
	\href {\doibase 10.1364/CQO.2013.M6.35} {\emph {\bibinfo {booktitle} {The
				Rochester Conferences on Coherence and Quantum Optics and the Quantum
				Information and Measurement meeting}}}\ (\bibinfo  {publisher} {Optical
		Society of America},\ \bibinfo {year} {2013})\ p.\ \bibinfo {pages}
	{M6.35}\BibitemShut {NoStop}%
	\bibitem [{\citenamefont {Maynard}\ \emph {et~al.}(2014)\citenamefont
		{Maynard}, \citenamefont {Bretenaker},\ and\ \citenamefont
		{Goldfarb}}]{MBG14}%
	\BibitemOpen
	\bibfield  {author} {\bibinfo {author} {\bibfnamefont {M.-A.}\ \bibnamefont
			{Maynard}}, \bibinfo {author} {\bibfnamefont {F.}~\bibnamefont {Bretenaker}},
		\ and\ \bibinfo {author} {\bibfnamefont {F.}~\bibnamefont {Goldfarb}},\
	}\href {\doibase 10.1103/PhysRevA.90.061801} {\bibfield  {journal} {\bibinfo
		{journal} {Phys. Rev. A}\ }\textbf {\bibinfo {volume} {90}},\ \bibinfo
	{pages} {061801} (\bibinfo {year} {2014})}\BibitemShut {NoStop}%
\bibitem [{\citenamefont {de~Almeida}\ \emph {et~al.}(2014)\citenamefont
	{de~Almeida}, \citenamefont {Sales}, \citenamefont {Maynard}, \citenamefont
	{Laupr\^etre}, \citenamefont {Bretenaker}, \citenamefont {Felinto},
	\citenamefont {Goldfarb},\ and\ \citenamefont {Tabosa}}]{ASM14}%
\BibitemOpen
\bibfield  {author} {\bibinfo {author} {\bibfnamefont {A.~J.~F.}\
		\bibnamefont {de~Almeida}}, \bibinfo {author} {\bibfnamefont
		{J.}~\bibnamefont {Sales}}, \bibinfo {author} {\bibfnamefont {M.-A.}\
		\bibnamefont {Maynard}}, \bibinfo {author} {\bibfnamefont {T.}~\bibnamefont
		{Laupr\^etre}}, \bibinfo {author} {\bibfnamefont {F.}~\bibnamefont
		{Bretenaker}}, \bibinfo {author} {\bibfnamefont {D.}~\bibnamefont {Felinto}},
	\bibinfo {author} {\bibfnamefont {F.}~\bibnamefont {Goldfarb}}, \ and\
	\bibinfo {author} {\bibfnamefont {J.~W.~R.}\ \bibnamefont {Tabosa}},\ }\href
{\doibase 10.1103/PhysRevA.90.043803} {\bibfield  {journal} {\bibinfo
		{journal} {Phys. Rev. A}\ }\textbf {\bibinfo {volume} {90}},\ \bibinfo
	{pages} {043803} (\bibinfo {year} {2014})}\BibitemShut {NoStop}%
\bibitem [{\citenamefont {Neveu}\ \emph {et~al.}(2017)\citenamefont {Neveu},
	\citenamefont {Maynard}, \citenamefont {Bouchez}, \citenamefont {Lugani},
	\citenamefont {Ghosh}, \citenamefont {Bretenaker}, \citenamefont {Goldfarb},\
	and\ \citenamefont {Brion}}]{NMB17}%
\BibitemOpen
\bibfield  {author} {\bibinfo {author} {\bibfnamefont {P.}~\bibnamefont
		{Neveu}}, \bibinfo {author} {\bibfnamefont {M.-A.}\ \bibnamefont {Maynard}},
	\bibinfo {author} {\bibfnamefont {R.}~\bibnamefont {Bouchez}}, \bibinfo
	{author} {\bibfnamefont {J.}~\bibnamefont {Lugani}}, \bibinfo {author}
	{\bibfnamefont {R.}~\bibnamefont {Ghosh}}, \bibinfo {author} {\bibfnamefont
		{F.}~\bibnamefont {Bretenaker}}, \bibinfo {author} {\bibfnamefont
		{F.}~\bibnamefont {Goldfarb}}, \ and\ \bibinfo {author} {\bibfnamefont
		{E.}~\bibnamefont {Brion}},\ }\href {\doibase 10.1103/PhysRevLett.118.073605}
{\bibfield  {journal} {\bibinfo  {journal} {Phys. Rev. Lett.}\ }\textbf
	{\bibinfo {volume} {118}},\ \bibinfo {pages} {073605} (\bibinfo {year}
	{2017})}\BibitemShut {NoStop}%
\bibitem [{\citenamefont {Eilam}\ \emph {et~al.}(2010)\citenamefont {Eilam},
	\citenamefont {Azuri}, \citenamefont {Sharypov}, \citenamefont
	{Wilson-Gordon},\ and\ \citenamefont {Friedmann}}]{EAS10}%
\BibitemOpen
\bibfield  {author} {\bibinfo {author} {\bibfnamefont {A.}~\bibnamefont
		{Eilam}}, \bibinfo {author} {\bibfnamefont {I.}~\bibnamefont {Azuri}},
	\bibinfo {author} {\bibfnamefont {A.~V.}\ \bibnamefont {Sharypov}}, \bibinfo
	{author} {\bibfnamefont {A.~D.}\ \bibnamefont {Wilson-Gordon}}, \ and\
	\bibinfo {author} {\bibfnamefont {H.}~\bibnamefont {Friedmann}},\ }\href
{\doibase 10.1364/OL.35.000772} {\bibfield  {journal} {\bibinfo  {journal}
		{Opt. Lett.}\ }\textbf {\bibinfo {volume} {35}},\ \bibinfo {pages} {772}
	(\bibinfo {year} {2010})}\BibitemShut {NoStop}%
\bibitem [{\citenamefont {Arimondo}(1996)}]{ARI96}%
\BibitemOpen
\bibfield  {author} {\bibinfo {author} {\bibfnamefont {E.}~\bibnamefont
		{Arimondo}},\ }\href {\doibase
	http://dx.doi.org/10.1016/S0079-6638(08)70531-6} {\bibfield  {journal}
	{\bibinfo  {journal} {Prog. Optics}\ }\textbf {\bibinfo {volume} {35}},\
	\bibinfo {pages} {257 } (\bibinfo {year} {1996})}\BibitemShut {NoStop}%
\bibitem [{\citenamefont {Singh}\ and\ \citenamefont
	{Natarajan}(2015)}]{SIN15}%
\BibitemOpen
\bibfield  {author} {\bibinfo {author} {\bibfnamefont {A.~K.}\ \bibnamefont
		{Singh}}\ and\ \bibinfo {author} {\bibfnamefont {V.}~\bibnamefont
		{Natarajan}},\ }\href {http://stacks.iop.org/1367-2630/17/i=3/a=033044}
{\bibfield  {journal} {\bibinfo  {journal} {New J. Phys.}\ }\textbf {\bibinfo
		{volume} {17}},\ \bibinfo {pages} {033044} (\bibinfo {year}
	{2015})}\BibitemShut {NoStop}%
\bibitem [{\citenamefont {Vafafard}\ \emph {et~al.}(2016)\citenamefont
	{Vafafard}, \citenamefont {Mahmoudi},\ and\ \citenamefont {Agarwal}}]{VMA16}%
\BibitemOpen
\bibfield  {author} {\bibinfo {author} {\bibfnamefont {A.}~\bibnamefont
		{Vafafard}}, \bibinfo {author} {\bibfnamefont {M.}~\bibnamefont {Mahmoudi}},
	\ and\ \bibinfo {author} {\bibfnamefont {G.~S.}\ \bibnamefont {Agarwal}},\
}\href {\doibase 10.1103/PhysRevA.93.033848} {\bibfield  {journal} {\bibinfo
	{journal} {Phys. Rev. A}\ }\textbf {\bibinfo {volume} {93}},\ \bibinfo
{pages} {033848} (\bibinfo {year} {2016})}\BibitemShut {NoStop}%
\bibitem [{\citenamefont {Pandey}\ \emph {et~al.}(2009)\citenamefont {Pandey},
	\citenamefont {Singh}, \citenamefont {Kumar}, \citenamefont {Suryanarayana},\
	and\ \citenamefont {Natarajan}}]{PSK09}%
\BibitemOpen
\bibfield  {author} {\bibinfo {author} {\bibfnamefont {K.}~\bibnamefont
		{Pandey}}, \bibinfo {author} {\bibfnamefont {A.~K.}\ \bibnamefont {Singh}},
	\bibinfo {author} {\bibfnamefont {P.~V.~K.}\ \bibnamefont {Kumar}}, \bibinfo
	{author} {\bibfnamefont {M.~V.}\ \bibnamefont {Suryanarayana}}, \ and\
	\bibinfo {author} {\bibfnamefont {V.}~\bibnamefont {Natarajan}},\ }\href
{\doibase 10.1103/PhysRevA.80.022518} {\bibfield  {journal} {\bibinfo
		{journal} {Phys. Rev. A}\ }\textbf {\bibinfo {volume} {80}},\ \bibinfo
	{pages} {022518} (\bibinfo {year} {2009})}\BibitemShut {NoStop}%
\bibitem [{\citenamefont {Singh}\ \emph {et~al.}(2013)\citenamefont {Singh},
	\citenamefont {Angom},\ and\ \citenamefont {Natarajan}}]{SAN13}%
\BibitemOpen
\bibfield  {author} {\bibinfo {author} {\bibfnamefont {A.~K.}\ \bibnamefont
		{Singh}}, \bibinfo {author} {\bibfnamefont {D.}~\bibnamefont {Angom}}, \ and\
	\bibinfo {author} {\bibfnamefont {V.}~\bibnamefont {Natarajan}},\ }\href
{\doibase 10.1103/PhysRevA.87.012512} {\bibfield  {journal} {\bibinfo
		{journal} {Phys. Rev. A}\ }\textbf {\bibinfo {volume} {87}},\ \bibinfo
	{pages} {012512} (\bibinfo {year} {2013})}\BibitemShut {NoStop}%
\bibitem [{\citenamefont {{T. Yang}}\ \emph {et~al.}(2015)\citenamefont {{T.
			Yang}}, \citenamefont {{K. Pandey}}, \citenamefont {{M. S. Pramod}},
	\citenamefont {{F. Leroux}}, \citenamefont {{C. C. Kwong}}, \citenamefont
	{{E. Hajiyev}}, \citenamefont {{Z. Y. Chia}}, \citenamefont {{B. Fang}},\
	and\ \citenamefont {{D. Wilkowski}}}]{YPP15}%
\BibitemOpen
\bibfield  {author} {\bibinfo {author} {\bibnamefont {{T. Yang}}}, \bibinfo
	{author} {\bibnamefont {{K. Pandey}}}, \bibinfo {author} {\bibnamefont {{M.
				S. Pramod}}}, \bibinfo {author} {\bibnamefont {{F. Leroux}}}, \bibinfo
	{author} {\bibnamefont {{C. C. Kwong}}}, \bibinfo {author} {\bibnamefont {{E.
				Hajiyev}}}, \bibinfo {author} {\bibnamefont {{Z. Y. Chia}}}, \bibinfo
	{author} {\bibnamefont {{B. Fang}}}, \ and\ \bibinfo {author} {\bibnamefont
		{{D. Wilkowski}}},\ }\href {\doibase 10.1140/epjd/e2015-60288-y} {\bibfield
	{journal} {\bibinfo  {journal} {Eur. Phys. J. D}\ }\textbf {\bibinfo {volume}
		{69}},\ \bibinfo {pages} {226} (\bibinfo {year} {2015})}\BibitemShut
{NoStop}%
\bibitem [{\citenamefont {Khan}\ \emph {et~al.}(2017)\citenamefont {Khan},
	\citenamefont {Kumar}, \citenamefont {Bharti},\ and\ \citenamefont
	{Natarajan}}]{KKB17}%
\BibitemOpen
\bibfield  {author} {\bibinfo {author} {\bibfnamefont {S.}~\bibnamefont
		{Khan}}, \bibinfo {author} {\bibfnamefont {M.~P.}\ \bibnamefont {Kumar}},
	\bibinfo {author} {\bibfnamefont {V.}~\bibnamefont {Bharti}}, \ and\ \bibinfo
	{author} {\bibfnamefont {V.}~\bibnamefont {Natarajan}},\ }\href {\doibase
	10.1140/epjd/e2017-70676-x} {\bibfield  {journal} {\bibinfo  {journal} {The
			European Physical Journal D}\ }\textbf {\bibinfo {volume} {71}},\ \bibinfo
	{pages} {38} (\bibinfo {year} {2017})}\BibitemShut {NoStop}%
\bibitem [{\citenamefont {Thomas}\ and\ \citenamefont {Quivers}(1980)}]{THQ80}%
\BibitemOpen
\bibfield  {author} {\bibinfo {author} {\bibfnamefont {J.~E.}\ \bibnamefont
		{Thomas}}\ and\ \bibinfo {author} {\bibfnamefont {W.~W.}\ \bibnamefont
		{Quivers}},\ }\href {\doibase 10.1103/PhysRevA.22.2115} {\bibfield  {journal}
	{\bibinfo  {journal} {Phys. Rev. A}\ }\textbf {\bibinfo {volume} {22}},\
	\bibinfo {pages} {2115} (\bibinfo {year} {1980})}\BibitemShut {NoStop}%
\bibitem [{\citenamefont {Mallick}\ \emph {et~al.}(2017)\citenamefont
	{Mallick}, \citenamefont {Dey},\ and\ \citenamefont {Pandey}}]{MDP17}%
\BibitemOpen
\bibfield  {author} {\bibinfo {author} {\bibfnamefont {N.~S.}\ \bibnamefont
		{Mallick}}, \bibinfo {author} {\bibfnamefont {T.~N.}\ \bibnamefont {Dey}}, \
	and\ \bibinfo {author} {\bibfnamefont {K.}~\bibnamefont {Pandey}},\ }\href
{http://stacks.iop.org/0953-4075/50/i=19/a=195502} {\bibfield  {journal}
	{\bibinfo  {journal} {Journal of Physics B: Atomic, Molecular and Optical
			Physics}\ }\textbf {\bibinfo {volume} {50}},\ \bibinfo {pages} {195502}
	(\bibinfo {year} {2017})}\BibitemShut {NoStop}%
\bibitem [{\citenamefont {McFerran}(2016)}]{MCF16}%
\BibitemOpen
\bibfield  {author} {\bibinfo {author} {\bibfnamefont {J.~J.}\ \bibnamefont
		{McFerran}},\ }\href {\doibase 10.1364/JOSAB.33.001278} {\bibfield  {journal}
	{\bibinfo  {journal} {J. Opt. Soc. Am. B}\ }\textbf {\bibinfo {volume}
		{33}},\ \bibinfo {pages} {1278} (\bibinfo {year} {2016})}\BibitemShut
{NoStop}%
\end{thebibliography}


\end{document}